\documentclass[12pt]{iopart}

\usepackage{amssymb,color}
\usepackage{amstext}
\usepackage{graphicx}
\usepackage[subrefformat=parens,labelformat=parens]{subfig}
\newcommand{\ket}[2][]{{|#2\rangle_{#1}}}
\newcommand{\bra}[2][]{{}_{#1}\langle #2|}
\newcommand{\braket}[3][]{{{}_{#1}\langle#2|#3\rangle_{#1}}}
\newcommand{\proj}[2][]{\ket{#2}_{#1}\bra{#2}}

\begin{document}

\title{Noisy propagation of coherent states in a lossy Kerr medium}

\author{Ludwig Kunz\textsuperscript{1,2}, Matteo G. A. Paris\textsuperscript{3}, and Konrad Banaszek\textsuperscript{1,2}}

\address{\textsuperscript{1} Centre of New Technologies, University of Warsaw, Banacha 2c, 02-097 Warszawa, Poland}
\address{\textsuperscript{2} Faculty of Physics, University of Warsaw, Pasteura 5, 02-093 Warszawa, Poland}
\address{\textsuperscript{3} Quantum Technology Lab, Dipartimento di Fisica, Universit\`{a} degli Studi di Milano, via Celoria 16, I-20133 Milano, Italy}

\ead{l.kunz@cent.uw.edu.pl}

\begin{abstract}
We identify and discuss nonlinear phase noise arising in Kerr self-phase modulation of a coherent light pulse propagating through an attenuating medium with third-order nonlinearity in a dispersion-free setting. This phenomenon, accompanying the standard unitary Kerr transformation of the optical field, is described with high accuracy as Gaussian phase diffusion with parameters given by closed expressions in terms of the system properties.
We show that the irreversibility of the nonlinear phase noise ultimately limits the ability to transmit classical information in the phase variable over a lossy single-mode bosonic channel with Kerr-type nonlinearity. Our 
model can be also used to estimate the amount of squeezing attainable through self-phase modulation in a Kerr medium with distributed attenuation.
\end{abstract}

\section{Introduction}

Third-order nonlinearity in optical media makes the effective refractive index dependent on light intensity. One of the consequences of this phenomenon, known as the Kerr effect, is self-phase modulation which consists in an intensity-dependent contribution to the phase shift acquired by a light beam propagating trough the medium. In the single-mode description, such propagation can be described by a nonlinear oscillator with one degree of freedom \cite{Milburn1986,PhysRevA.39.4628,1464-4266-1-6-308}. The quantum version of this model exhibits a wealth of nonclassical effects, such as squeezing \cite{RosenbluhShelbyPRL1991} and generation of Schr\"{o}dinger cat states \cite{YurkeStolerPRL1986}. However, their experimental realisation requires suppression of the accompanying decoherence mechanisms with a prominent role played by photon loss. While it is feasible to experimentally produce squeezed states using third-order nonlinearity in optical fibres \cite{Schmitt1998,Werner1998,Fiorentino2002}, Schr\"{o}dinger cat states turn out to be much more elusive owing to higher nonlinearities necessary for generation and greater susceptibility to losses \cite{Buzek1995}.

In parallel, optical nonlinearities have been recognised as a key factor limiting the information capacity of fibre optical communication links \cite{Mitra2001,Wegener2004,Ellis2010,Temprana1445}. Up to date, most studies in this area have used the classical description of electromagnetic fields in terms of complex amplitudes governed by deterministic equations of motion. In this approach, noise appears as a result of signal amplification, averaging over signals transmitted through other channels in a multiplexed system, or the functioning of the detection stage. The performance of conventional optical detection is limited by the shot noise which is a manifestation of quantum fluctuations in measured electromagnetic fields \cite{KikuchiJLT2016}. However, if the propagation of electromagnetic fields is both nonlinear and lossy, quantum fluctuations in the optical medium undergo a non-unitary transformation which may result in additional noise that would not emerge in a classical description. This eventuality motivates a full quantum mechanical study of nonlinear effects in optical signal propagation.

\begin{figure}[h!tp]
	\centering
	\includegraphics[width=.8\linewidth]{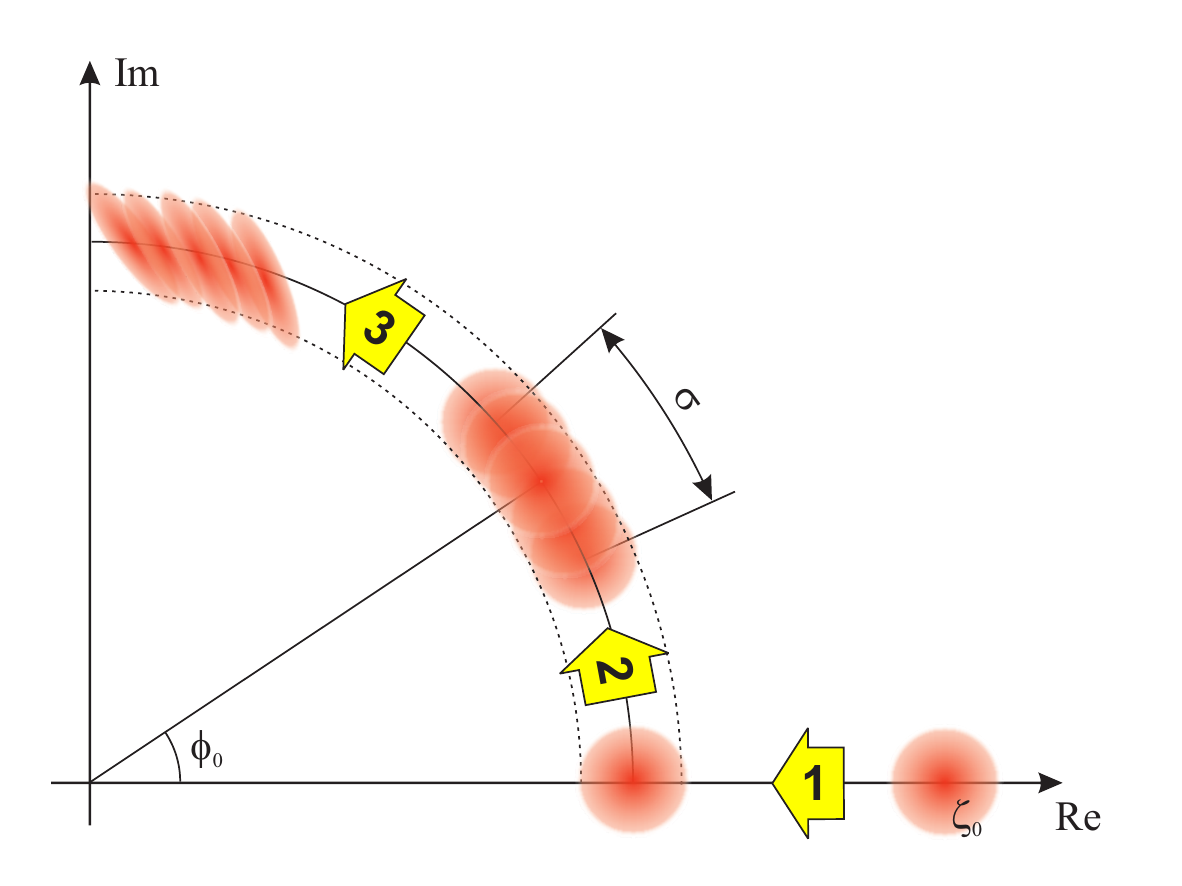}
	\caption{Decomposition of noisy propagation of a coherent light pulse in a lossy Kerr medium into a sequence of three processes. Process \textbf{1} is linear attenuation of the input complex amplitude $\zeta_0$ by a factor $\sqrt{\tau}$, where $\tau$ is the power transmission of the medium. Process \textbf{2} is nonlinear phase noise which shifts the phase by $\phi_0$ and introduces Gaussian phase diffusion characterised by variance $\sigma^2$. Finally, process \textbf{3} is unitary Kerr evolution that would have occurred in the absence of attenuation, applied to the output of process \textbf{2}.}
	\label{Fig:kerrdecomp}
\end{figure}

The purpose of the present paper is to analyse a single-mode model for self-phase modulation of a coherent light pulse propagating through a lossy nonlinear dispersion-free medium. 
Our central objective will be to identify and characterise quantitatively excess noise which occurs as a result of nonlinear transformation of quantum fluctuations in the course of lossy propagation and cannot be compensated at the output with the help of a reversible unitary transformation. Specifically, we show that lossy Kerr propagation can be mathematically decomposed into a sequence of three distinct processes that are formally applied to the input coherent state one after another. This decomposition is represented pictorially in Fig.~\ref{Fig:kerrdecomp}. The first process is standard attenuation which lowers the complex amplitude of the input coherent state. The last one is unitary Kerr evolution which would have occurred in the absence of loss. The intermediary process we will focus our attention on arises solely a result of a non-trivial interplay between the Kerr nonlinearity and the optical loss. We demonstrate that it has the form of nonlinear phase noise which can be approximated with high accuracy by a Gaussian distribution for the additive random phase. We provide closed analytical expressions for parameters that characterise the properties of this decoherence mechanism.

The derived effective description of the nonlinear phase noise is employed to discuss its impact on the transmission of classical information in the phase variable. We choose continuous phase modulation with constant intensity as the encoding format \cite{KahnHoSTQ2004} and characterise accessible information using the Holevo quantity  \cite{HolevoPIT1973,HolevoTIT1998,SchumacherWestmorelandPRA1997}, which takes into account the most general quantum mechanical measurement strategies at the channel output, including compensation of reversible nonlinear effects. It is shown that for the propagation over a lossy Kerr medium the Holevo quantity exhibits a maximum in the signal intensity and that the nonlinear phase noise renders the phase variable useless for information encoding at high signal powers. Consequently, this effect can be expected to alter substantially the ultimate communication capacity of an optical channel compared to the linear case, which has been analysed rigorously in recent years \cite{GiovannettiGarciaPatronNPH2014}. As another application of the presented description we study generation of squeezing using the Kerr effect and derive a simple estimate for the impact of distributed loss in the squeezing medium on the output quadrature variance.

This paper is organised as follows. In Sec.~\ref{Sec:model} we introduce the theoretical model for the nonlinear propagation. In Sec.~\ref{Sec:phase_noise} we derive the effective description for the excess noise generated in the model. Based on this description we demonstrate in Sec.~\ref{Sec:Holevo} how the excess noise impacts the accessible information for the phase variable. In Sec.~\ref{Sec:squeezing} we discuss the generation of squeezed states using a lossy Kerr medium and derive a simple estimate for the attainable reduction of quadrature noise in the case of distributed attenuation. Finally, Sec.~\ref{Sec:Conclusions} closes the paper with some
concluding remarks.

\section{Lossy nonlinear propagation}
\label{Sec:model}

In a single-mode model describing dispersion-free propagation of an optical pulse in a lossy Kerr medium along a direction $z$, evolution of the system is given by the Master equation
\begin{equation}
\frac{d\hat{\varrho}}{dz} = i \mu [\hat{n}^2, \hat{\varrho}(z)] + \frac{\alpha}{2}
\left( 2 \hat{a} \hat{\varrho}(z) \hat{a}^\dagger - \hat{n}\hat{\varrho}(z) - \hat{\varrho}(z)\hat{n}\right)
\end{equation}
with the mode creation and annihilation operators $\hat{a}^\dagger$ and $\hat{a}$ respectively, the photon number operator $\hat{n} = \hat{a}^\dagger\hat{a}$, the nonlinearity strength $\mu$ and the parameter $\alpha$ characterising linear losses.
In order to separate reversible phenomena, we will carry out the analysis in the interaction picture defined by the nonlinear Hamiltonian $\mu \hat{n}^2$ of the Kerr interaction and represent the field density operator $\hat{\varrho}(z)$ as
\begin{equation}
\hat{\varrho}(z) = \exp(i \mu z \hat{n}^2) \hat{\varrho}'(z) \exp(- i \mu z \hat{n}^2).
\label{Eq:rhorho'}
\end{equation}
One should note the opposite sign in the exponent of the evolution operator compared to the standard Heisenberg picture in the time domain, as the formalism is now applied to propagation along a spatial dimension.
The Master equation for the transformed density operator $\hat\varrho'(z)$ reads
\begin{equation}
\frac{d\hat{\varrho}'}{dz} = \frac{\alpha}{2} \left( 2 \hat{a}'(z) \hat{\varrho}'(z)
[\hat{a}'(z)]^\dagger - \hat{n} \hat{\varrho}'(z) - \hat{\varrho}'(z) \hat{n} \right)
\label{Eq:drho'dt}
\end{equation}
with the time-dependent annihilation operator in the interaction picture $\hat{a}'(z) = \exp(-i \mu z \hat{n}^2) \hat{a} \exp(i \mu z \hat{n}^2)$ and the photon number operator $\hat{n}$ invariant with respect to the transformation. We assume that the input state is a coherent state with a complex amplitude $\zeta_0$, $\hat{\varrho}(0) = \proj{\zeta_0}$, given in the Fock basis by a superposition
\begin{equation}
\ket{\zeta_0} = e^{-\bar n/2} \sum_{n=0}^{\infty} \frac{\zeta_0^n}{\sqrt{n!}} \ket{n} .
\end{equation}
We will use $\bar n = |\zeta_0|^2$ to denote the mean photon number in the input coherent state.
The propagation equations for the above model can be solved analytically \cite{PhysRevA.39.4628,1464-4266-1-6-308}.
In order to determine the elements of the field density matrix after a propagation distance $z$ in the Fock basis
${\hat{\varrho}_{mn}'(z) = \bra{m} \hat{\varrho}'(z) \ket{n}}$,
one can use an ansatz
\begin{equation}
\varrho'_{mn}(z) = \frac{\zeta_0^m (\zeta_0^\ast)^n}{\sqrt{m! n!}} e^{-(m+n)\alpha z/2} c_{m-n}(z) .
\label{Eq:ansatz}
\end{equation}
The factor $c_{j}(z)$ depends only on the difference $j=m-n$ between the matrix element indices. It is easy to verify that the above ansatz, after inserting into Eq.~(\ref{Eq:drho'dt}), yields a closed first-order differential equation for $c_j(z)$ which can be solved by standard means. For completeness, we present the detailed derivation in \ref{Sec:deriv}.

It will be convenient to characterise the medium with two dimensionless parameters. The first one is the transmission $\tau=\exp(-\alpha z)$ that specifies the fraction of power 
$\tau \bar n$
retained in the field at the channel output. The second one is the ratio $\varkappa = \mu/\alpha$ describing the strength of the nonlinear interaction with respect to attenuation. Note that $\varkappa$ is an intensive parameter characterising the bulk medium itself, while $\tau$ depends on the channel length. Using this parameterisation, the solution of the Master equation can be written compactly as
\begin{equation}
\varrho'_{mn}(z) = \braket{m}{\sqrt{\tau}\zeta_0} \braket{\sqrt{\tau}\zeta_0}{n} \exp[-\bar n f_\tau((m-n)\varkappa)] .
\label{Eq:density_elements}
\end{equation}
The product of the first two terms $\braket{m}{\sqrt{\tau}\zeta_0} \braket{\sqrt{\tau}\zeta_0}{n}$ describes a coherent state with an attenuated amplitude $\sqrt{\tau}\zeta_0$. Because unitary Kerr evolution has been included in the transformation to the interaction picture in Eq.~(\ref{Eq:rhorho'}), the last factor $\exp[-\bar n f_\tau((m-n)\varkappa)]$ appearing on the right hand side of Eq.~(\ref{Eq:density_elements}) only contains effects arising from the combination of the Kerr nonlinearity and loss. The complex function in the exponent is given explicitly by
\begin{equation}
f_\tau(\varkappa) = 1-\tau - \frac{1- \tau^{1-2i\varkappa } }{1 - 2i\varkappa}.
\label{Eq:f}
\end{equation}

\begin{table}
\begin{center}
\begin{tabular}{cccccc}
\hline
Fibre & Wavelength & Attenuation  & Nonlinearity & Photon energy  & Pulse duration \\
type & [nm] & $\alpha$ $[\text{km}^{-1}]$ & $\gamma_{\text{NL}}$ $[\text{km}^{-1} \text{W}^{-1}]$ & $\hbar\omega_0$ [J] &  $T$ [fs] \\
\hline
SMF-28 & 1310 & 0.074 & $1$ & $15.2 \times 10^{-20}$& 410 \\
SMF-28 & 1550 & 0.046 & $1$ & $12.8 \times 10^{-20}$ & 560 \\
HB1500 & 1550 & 0.46 & $3$ & $12.8 \times 10^{-20}$ & 170\\
\hline
\end{tabular}
\end{center}
\caption{Exemplary combinations of optical fibre and pulse parameters that give the value of the dimensionless nonlinearity parameter $\varkappa = 5 \times 10^{-6}$.}
\label{Tab:parameters}
\end{table}

\begin{figure}
	\centering
	\subfloat[$\tau\bar n=1$\label{Fig:Q0}]{\includegraphics[width=.5\linewidth]{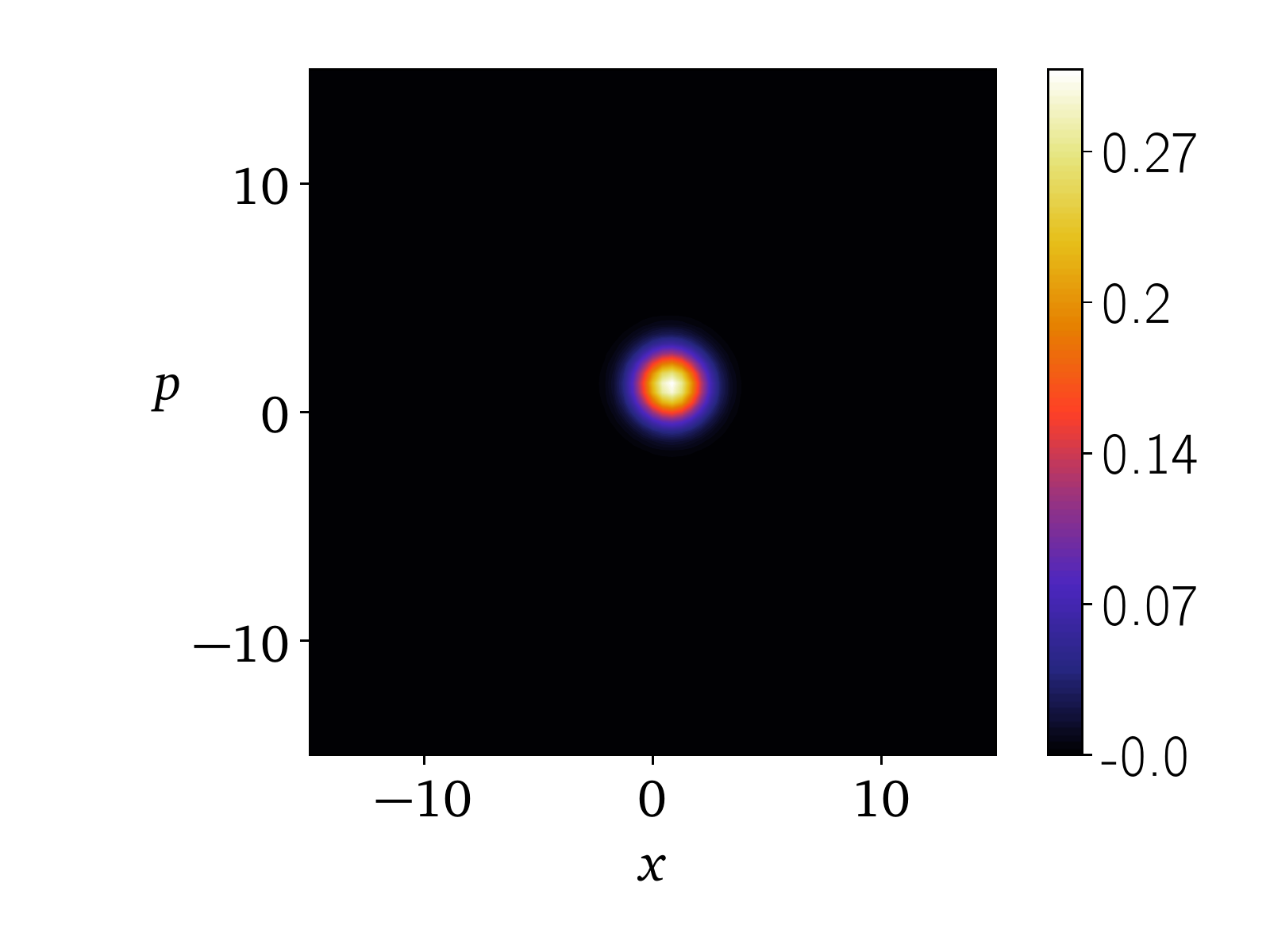}}
	\subfloat[$\tau\bar n=3$\label{Fig:Q1}]{\includegraphics[width=.5\linewidth]{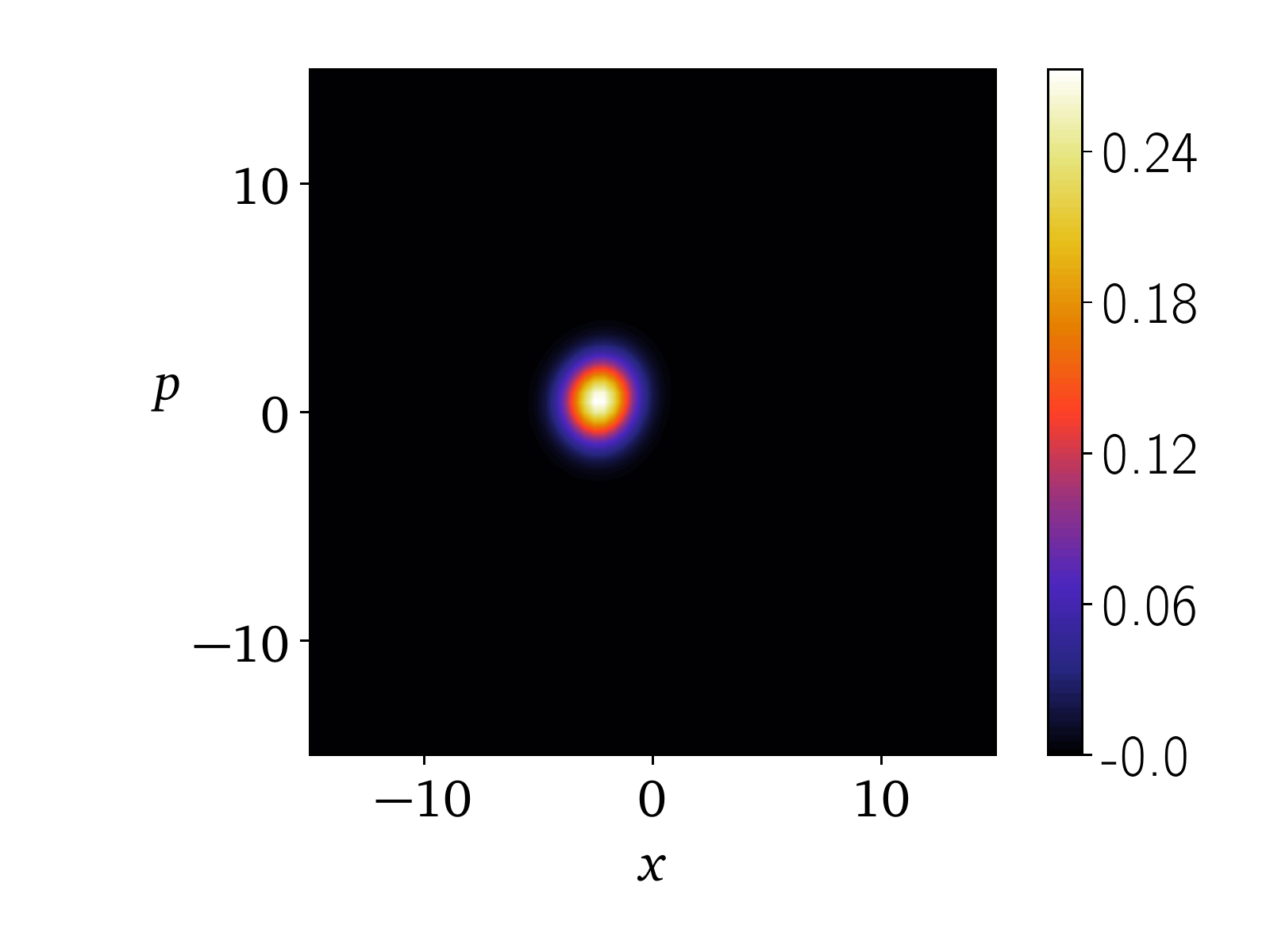}} \\
	\subfloat[$\tau\bar n=15$\label{Fig:Q2}]{\includegraphics[width=.5\linewidth]{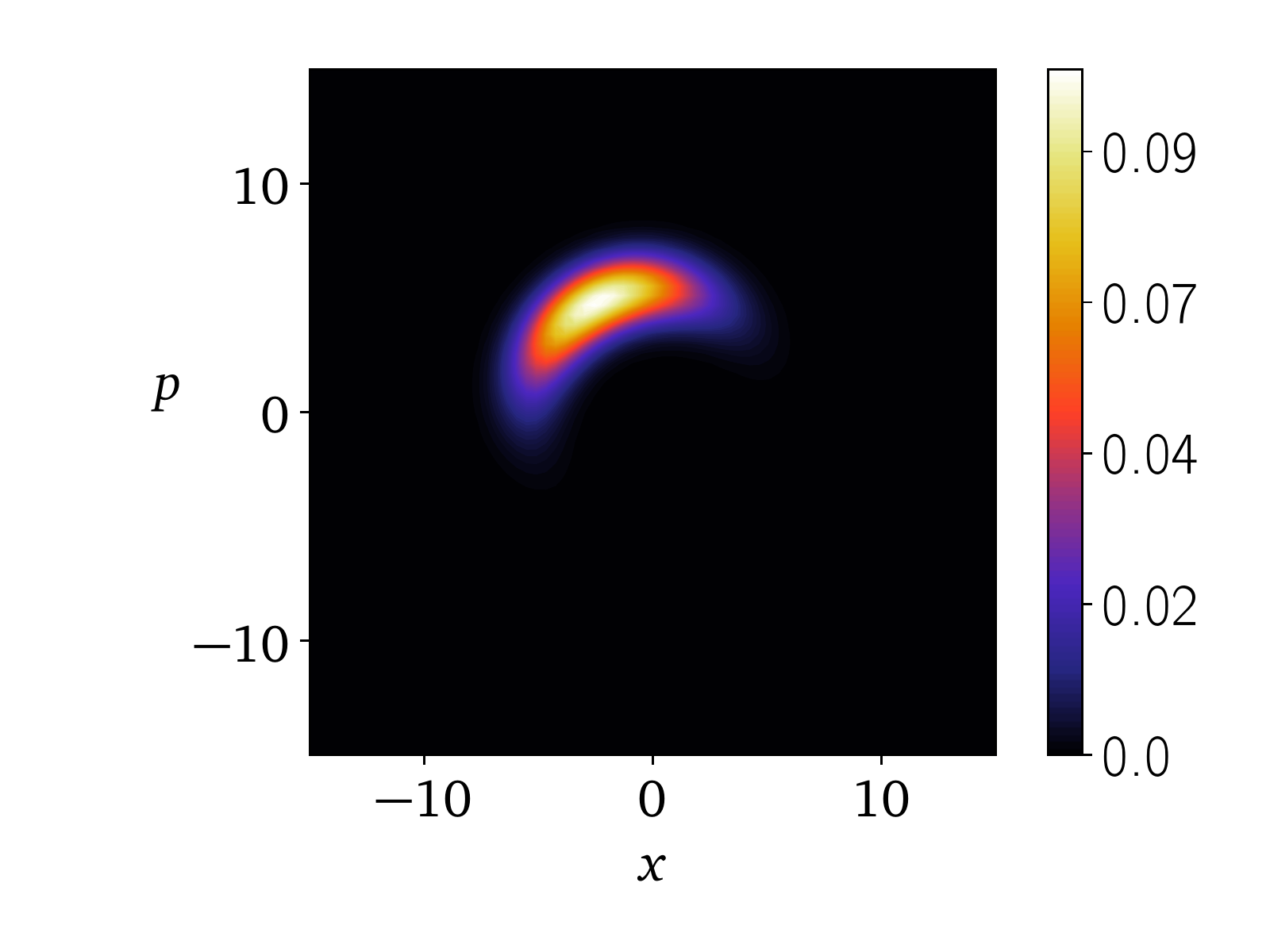}}
	\subfloat[$\tau\bar n=60$\label{Fig:Q3}]{\includegraphics[width=.5\linewidth]{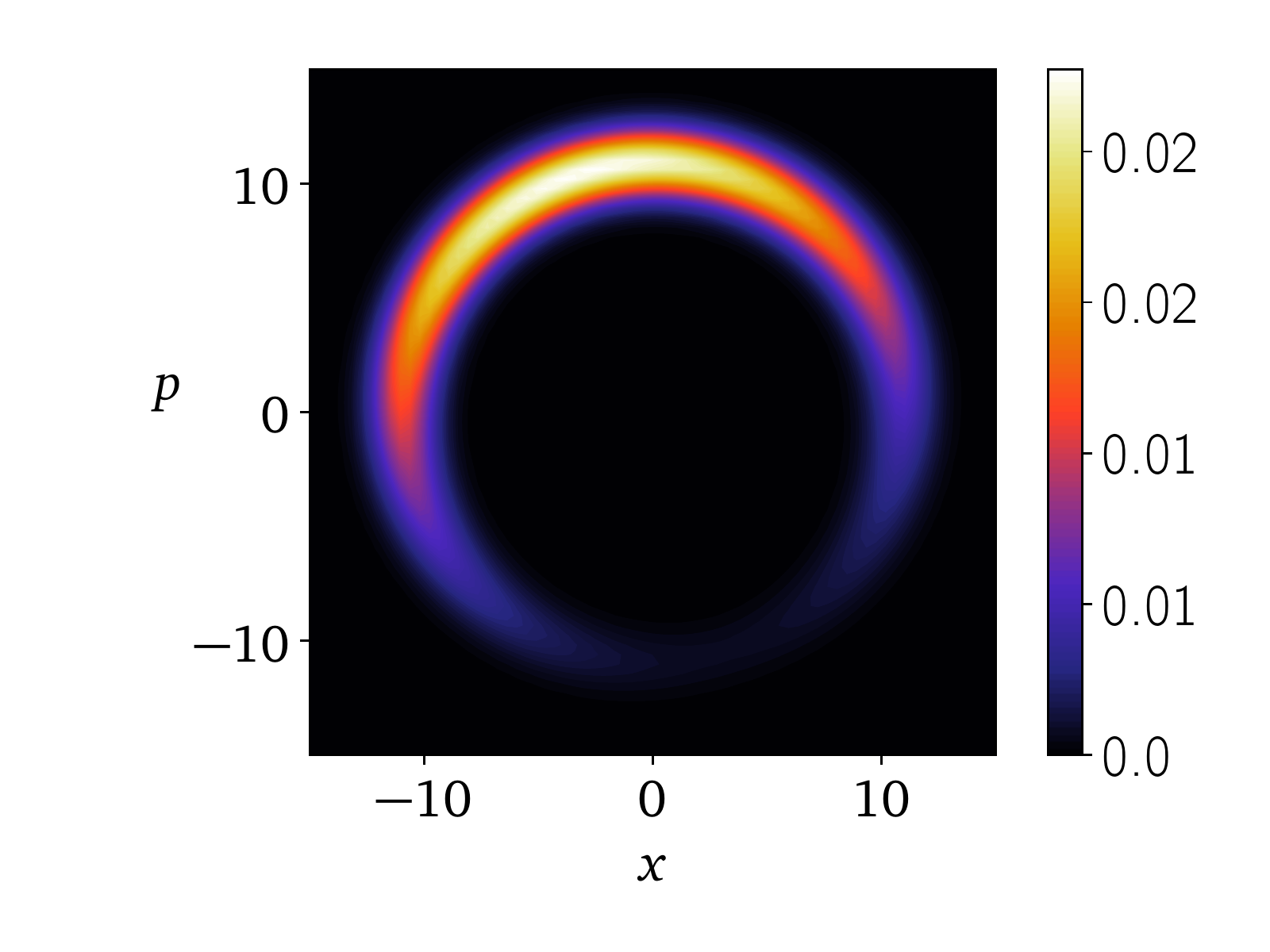}}
	\caption{The Husimi $Q$ function of the output state emerging from a lossy nonlinear medium with transmission $\tau = 10^{-8}$ and nonlinearity $\varkappa = 5 \times 10^{-6}$, shown for several different output mean photon numbers $\tau \bar n$. A coherent state with a real amplitude $\sqrt{\bar n}$ has been assumed at the input. Nonlinear phase noise is clearly seen to increase with the pulse intensity.}
	\label{Fig:state_evolution}
\end{figure}

In Fig.~\ref{Fig:state_evolution} we depict the Husimi $Q$ function for the state $\hat{\varrho}'(z)$ with an increasing mean photon number. The dimensioneless nonlinearity for the example had been chosen as $\varkappa = 5 \times 10^{-6}$. It can be related to parameters of actual fibre optic links as $\varkappa = \gamma_\text{NL} \hbar\omega_0/(\alpha T)$, where $\gamma_\text{NL}$ is the fibre third-order nonlinearity, $\hbar\omega_0$ is the energy of a single photon at the carier frequency $\omega_0$, and $T$ is the pulse duration. Combinations of these parameters that give the examplary value of $\varkappa$ for two common types of optical fibre have been collected in Table~\ref{Tab:parameters}. The channel transmission has been taken $\tau = 10^{-8}$, which corresponds to 400~km of a standard SMF-28 fibre. Two effects are clearly seen: the overall phase of the state is shifted and the state becomes spread on a circle. The spreading becomes stronger with increasing pulse intensity. It should be emphasised that this spreading does not arise from the reversible deformation of quadrature fluctuations by the unitary Kerr self-modulation. These effects are excluded in the density matrix in the interaction picture $\hat{\varrho}'(z)$ and furthermore they can be estimated to be negligible in the presented numerical examples: for a nonlinear strength of $\varkappa = 5 \times 10^{-6}$, a channel transmission of $\tau = 10^{-8}$ and a mean photon number at the output of $\tau \bar n = 60$ at the input, the relative change in quadrature fluctuations calculated using the approach presented in Sec.~\ref{Sec:squeezing} can be characterised using the squeezing parameter $r$ to remain below $2 r \approx 2.2 \times 10^{-2}$.

\section{Nonlinear phase noise}
\label{Sec:phase_noise}

\begin{figure}
	\centering
	\subfloat[ $\tau = 10^{-8}$\label{Fig:f1e-8}]{\includegraphics[width=.5\linewidth]{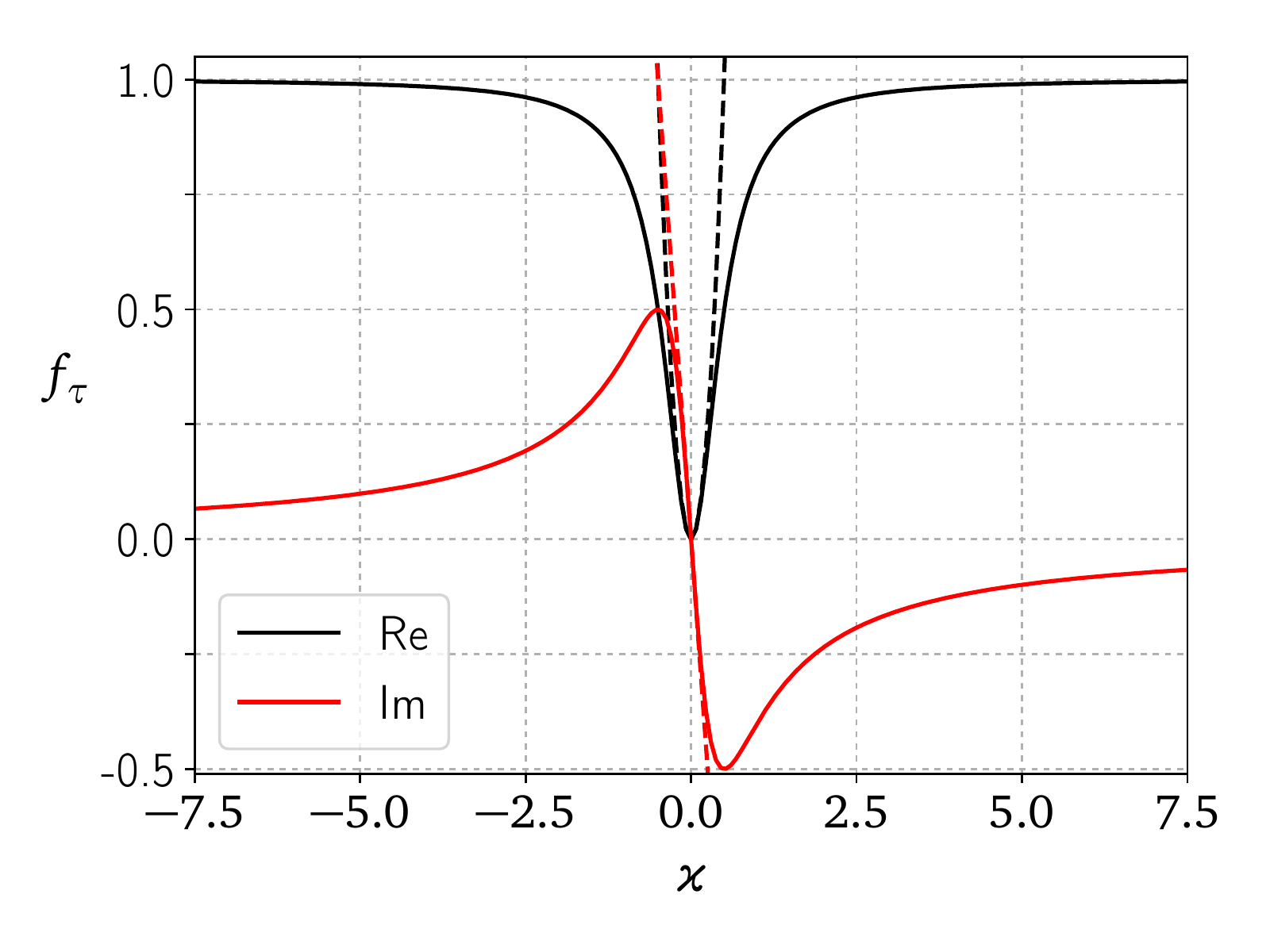}}
	\subfloat[$\tau = 0.8$\label{Fig:f08}]{\includegraphics[width=.5\linewidth]{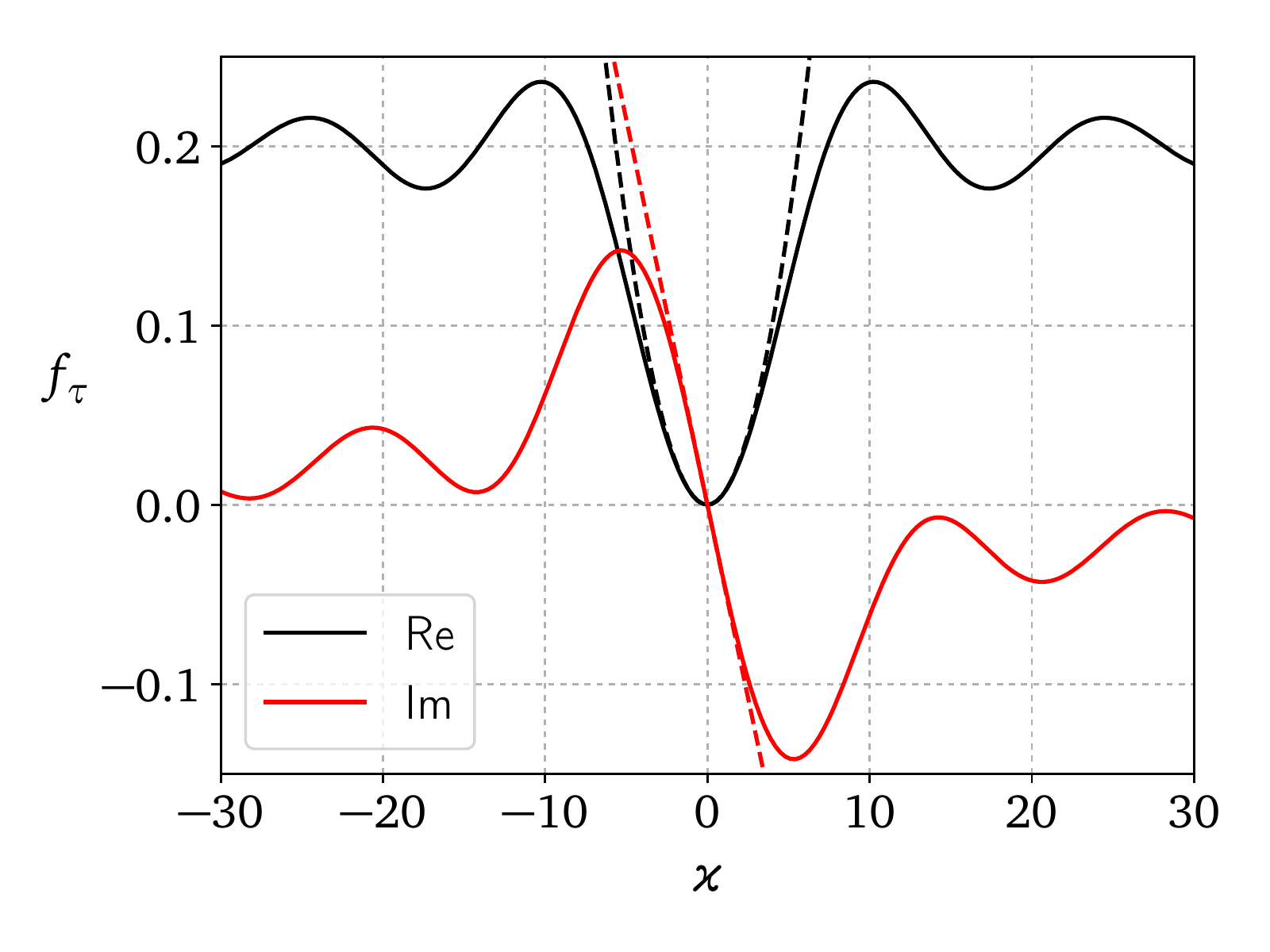}}
\caption{The real and the imaginary parts of the function $f_\tau(\varkappa)$ defined in Eq.~(\ref{Eq:f}) for two different transmissions $\tau$.}
\label{Fig:f}
\end{figure}

Let us now derive effective parameters that describe features seen in Fig.~\ref{Fig:state_evolution}. The starting point will be the inspection of the real and imaginary parts of the function $f_\tau(\varkappa)$ defined in Eq.~(\ref{Eq:f}), shown in Fig.~\ref{Fig:f} for transmissions $\tau=10^{-8}$ and $\tau=0.8$ as a function of the dimensionless nonlinearity $\varkappa$. As seen in Fig.~\subref*{Fig:f1e-8}, for $\tau \ll 1$ the real part $\text{Re}f_\tau(\varkappa)$ has the form of a dip centered around $\varkappa = 0$. Using Eq.~(\ref{Eq:f}), this dip can be modelled by an inverted Lorentzian proportional to $1/(1+4\varkappa^2)$, whose width is of the order of one. The relevant range of the argument of $f_\tau$ in Eq.~(\ref{Eq:density_elements}) is defined by the range of the Fock basis indices $m$ and $n$ for which the product of the scalar products $\braket{m}{\sqrt{\tau}\zeta_0} \braket{\sqrt{\tau}\zeta_0}{n}$ substantially differs from zero. Because $\bigl| \braket{m}{\sqrt{\tau}\zeta_0}\bigr|^2 $ is a Poissonian distribution in the integer variable $m$ with an average $\tau \bar n$ and hence standard deviation $\sqrt{\tau \bar n}$, this condition will be met for $|m-n| \lesssim\sqrt{\tau \bar n}$. Consequently, the argument of the function $f_\tau$ has the order of magnitude $\varkappa |m-n| \lesssim \varkappa \sqrt{\tau \bar n}$. If the product $\varkappa \sqrt{\tau \bar n} \ll 1 $, which is the case for the numerical example shown in the preceding section, it is justified to expand the function $f_\tau(\varkappa)$ up to the second order in $\varkappa$, which gives:
\begin{equation}
f_\tau(\varkappa) \approx - 2 i (1 - \tau + \tau \log \tau) \varkappa + [4 - 2\tau - 2\tau(1-\log \tau)^2]\varkappa^2 .
\label{Eq:approx_zeta}
\end{equation}

After the above approximation, the exponent multiplying Fock basis elements of the coherent state density matrix in Eq.~(\ref{Eq:density_elements}) contains terms linear and quadratic in $m-n$ with respectively imaginary and real multiplicative constants. It is easily verified by a direct calculation that such a density operator can be equivalently written as a statistical mixture of coherent states with a fixed mean photon number and the phase averaged using a Gaussian distribution with a mean $\phi_0$ and a standard deviation $\sigma$:
\begin{eqnarray}
\lefteqn{ \sum_{m,n=0}^{\infty} \ket{m}\bra{n} \braket{m}{\zeta} \braket{\zeta}{n} \exp [ i (m-n) \phi_0 - (m-n)^2 \sigma^2 /2]} & & \nonumber \\
& = & \int_{-\infty}^{\infty} d \phi \frac{1}{\sqrt{2\pi \sigma^2}} e^{-(\phi-\phi_0)^2/2\sigma^2} \proj{e^{i\phi}\zeta}.
\label{Eq:GaussianNoiseDensityMatrix}
\end{eqnarray}
For clarity we have denoted here $\zeta= \sqrt{\tau}\zeta_0$.
Using the explicit expression for the Fock state density matrix elements from Eq.~(\ref{Eq:density_elements}) with the approximation given in Eq.~(\ref{Eq:approx_zeta}) enables us to identify the parameters of the Gaussian distribution as:
\begin{eqnarray}
\phi_0 & = &  2 \varkappa \bar n (1-\tau + \tau \log \tau) , \label{Eq:phi0approx}\\
\sigma^2 & = & 4 \varkappa^2 \bar n [2-\tau - \tau(1-\log\tau)^2] . \label{Eq:sigmaapprox}
\label{Eq:phi0sigma}
\end{eqnarray}
It is seen that the linear term in the expansion $(\ref{Eq:approx_zeta})$ generates a phase shift $\phi_0$ and the quadratic term is responsible for the phase noise characterised by the variance $\sigma^2$. While the phase shift $\phi_0$ can be pre-compensated at the channel input, the phase noise clearly affects the ability to encode information in the phase variable. We will discuss quantitatively this phenomenon in Sec.~\ref{Sec:Holevo}.

\begin{figure}
	\centering
	\includegraphics[width=.7\linewidth]{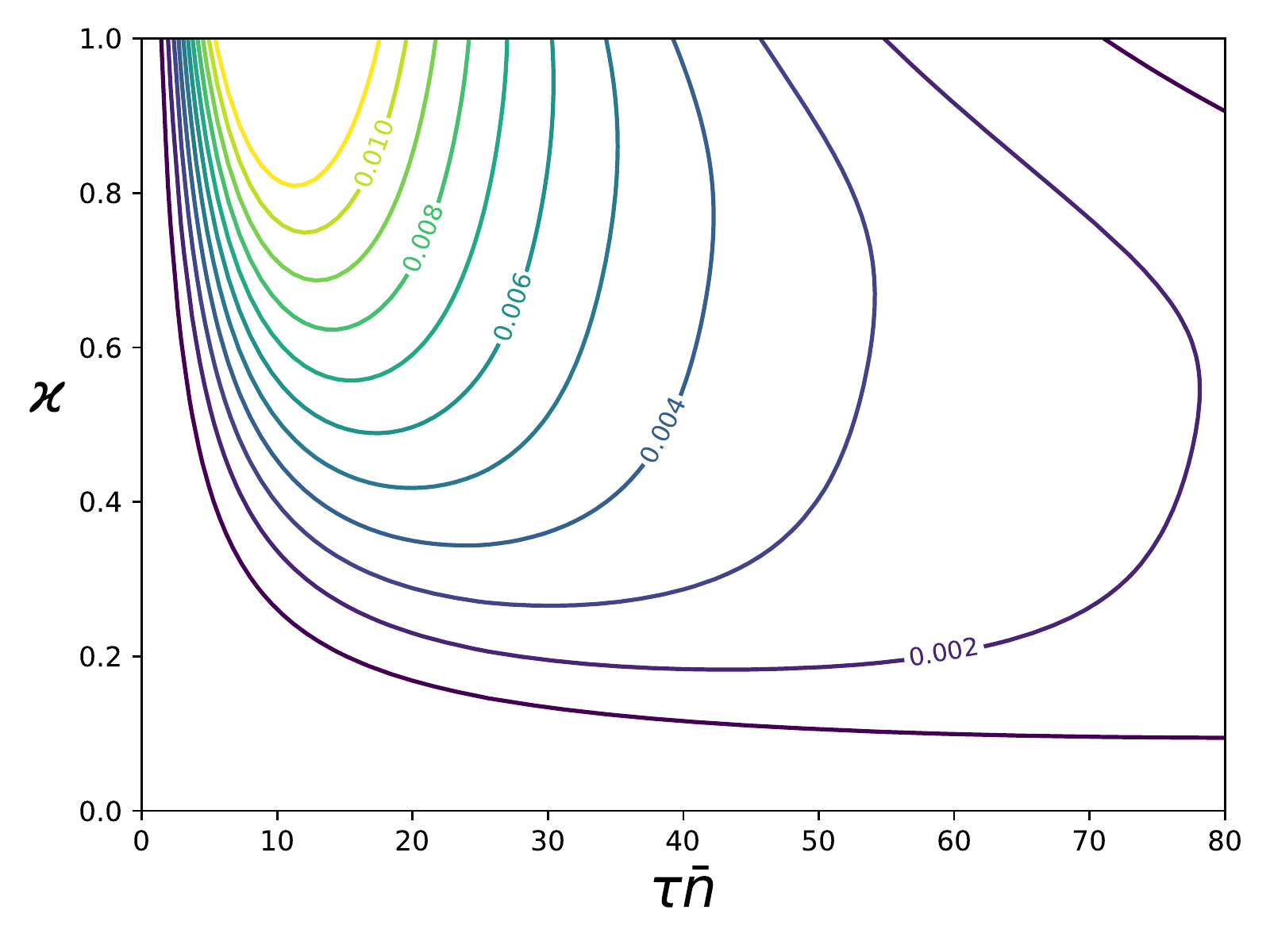}
	\caption{Infidelity $1-F$ between the actual density matrix $\hat{\varrho}'$ and its approximation $\hat{\varrho}_{G}'$ based on the Gaussian phase diffusion model for the transmission $\tau = 0.8$ as a function of the output mean photon number $\tau\bar{n}$ and the dimensionless nonlinearity $\varkappa$.}
	\label{Fig:Fidelity}
\end{figure}

As a side remark, let us note that the validity of the expansion (\ref{Eq:approx_zeta}) can be extended beyond the regime discussed above. This is because the function $f_\tau(\varkappa)$ appears in the exponent in Eq.~(\ref{Eq:density_elements}) multiplied by the input intensity $\bar n$. If this product is large compared to one, the exponential factor effectively suppresses the respective off-diagonal elements of the density matrix regardless of its specific form. In Fig.~\label{Fig:f08} we plot the real and imaginary parts of $f_\tau(\varkappa)$ for $\tau=0.8$ compared with the corresponding quadratic expansion. Because the asymptotic value for $|\varkappa| \gg 1 $ is $f_\tau(\varkappa) \approx 1-\tau$, the rule of thumb for the quadratic expansion to hold using the above reasoning is $(1-\tau)\bar n \gg 1$. The accuracy of this approximation can be assessed by comparing the infidelity $1-F = 1-\text{Tr} \sqrt{\sqrt{\hat{\varrho}'} \hat{\varrho}_{G}' \sqrt{\hat{\varrho}'}}$ between the actual density matrix $\hat{\varrho}'$ and its approximation $\hat{\varrho}_{G}'$ introduced in Eq.~(\ref{Eq:GaussianNoiseDensityMatrix}) with parameters of the Gaussian phase noise given in Eqs.~(\ref{Eq:phi0approx}) and (\ref{Eq:phi0sigma}). The infidelity is shown in Fig.~\ref{Fig:Fidelity} for $\tau=0.8$ as a function of the mean output photon number $\tau\bar n$ and the strength of the nonlinear interaction $\varkappa$. It is seen that the departure from the Gaussian approximation in the parameter space is localised in the region of high $\varkappa$ and moderate output mean photon numbers $\tau \bar n$, and works very well for higher pulse intensities. Notice that the upper limit for the range of $\varkappa$ in Fig.~\ref{Fig:Fidelity} has been taken well above realistic values for optical fibres. Notice also the
small values of the infidelity $1-F$, ensuring that the physical properties of the phase diffused state
are very close to those of the state subject to lossy nonlinear propagation \cite{fid1,fid2,fid3}.

\section{Phase shift keying}
\label{Sec:Holevo}

Nonlinear phase noise analysed in the preceding section is a non-deterministic phenomenon that cannot be compensated at the channel output in contrast to  unitary Kerr transformation. This affects the ability to transmit classical information in the phase variable which is routinely used in keying constellations for fibre optic communication, such as quadriphase shift keying \cite{Kazovsky2006}. Without phase noise, the amount of information that can be transmitted by modulating the phase of a light pulse grows with its intensity. Fixing intensity defines a ring in the complex amplitude plane which with increasing radius can accommodate more coherent states whose distinguishability is limited by the shot noise. This picture changes dramatically when channel nonlinearities induce phase noise. As seen in Eq.~(\ref{Eq:sigmaapprox}), the variance of the phase noise grows linearly with the input field intensity, which may actually obliterate information keyed in the phase.

\begin{figure}[htp]
	\centering
	\subfloat[\label{Fig:encoding}]{\includegraphics[width=.4\textwidth]{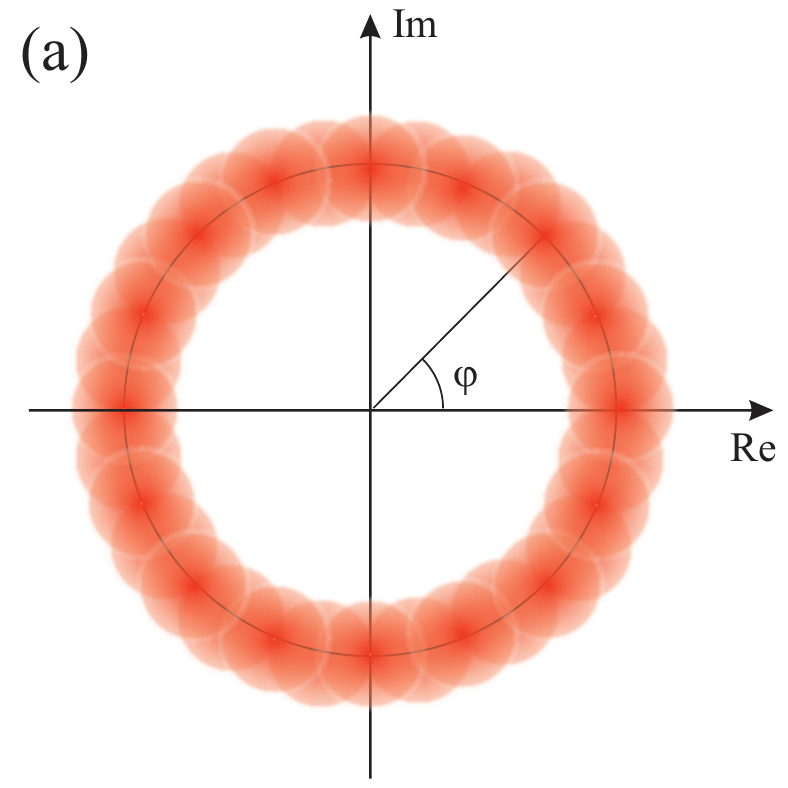}}
	\subfloat[\label{Fig:holevo}]{\includegraphics[width=.5\textwidth]{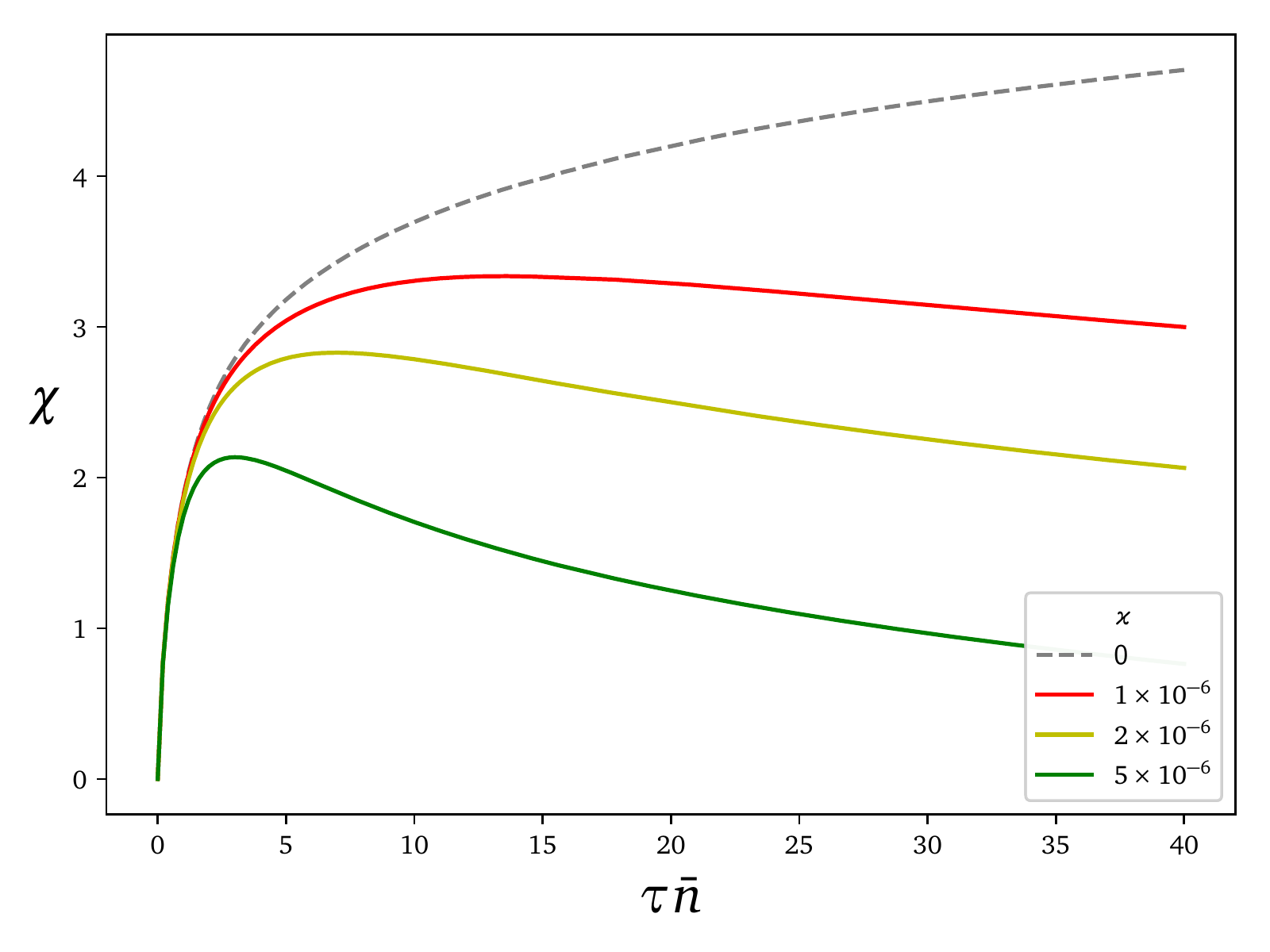}}
	\caption{(a) Continuous phase shift keying. Information is encoded in the phase $\varphi$ of a constellation of coherent states with constant intensity uniformly distributed on a circle. (b) Holevo quantity for the continuous phase shift keying constellation as a function of the output mean photon number $\tau\bar{n}$ for several values of the nonlinearity $\varkappa$.}
	\label{Fig:encoding_holevo}
\end{figure}

We will analyse quantitatively the above phenomenon by considering a keying constellation which consists of coherent states $\ket{e^{i\varphi}\zeta_0}$ continuously distributed on a circle with a uniform distribution $p_\varphi = 1/2\pi$ \cite{KahnHoSTQ2004}, as shown in Fig.~\subref*{Fig:encoding}. As a measure of accessible information we will take the Holevo quantity $\chi$, which takes into account the most general measurement strategies permitted in quantum mechanics \cite{HolevoPIT1973,HolevoTIT1998,SchumacherWestmorelandPRA1997}. The Holevo quantity is given as a difference between the von Neumann entropy $S(\hat \varrho) = - \text{Tr}\left(\hat \varrho \log \varrho\right)$ of the average state emerging from the channel and the average entropy of individual states and it is given in our case by
\begin{equation}
\chi = S \left(\int_{0}^{2\pi} d\varphi \, p_\varphi \hat{\varrho}'_\varphi \right) - \int_{0}^{2\pi} d\varphi \, p_\varphi S( \hat{\varrho}'_\varphi).
\label{Eq:Holevo}
\end{equation}
Because the von Neumann entropy is invariant with respect to unitary transformations we can exclude from our calculations the unitary Kerr process and take as $\hat{\varrho}'_\varphi$ phase diffused states given by Eq.~(\ref{Eq:GaussianNoiseDensityMatrix}) for $\zeta = \sqrt{\tau} e^{i\varphi}\zeta_0$. The phase averaged state appearing in the first term of Eq.~(\ref{Eq:Holevo}) as the argument of the von Neumann entropy  is a statistical mixture of Fock states with a Poissonian distribution characterised by the mean $\tau\bar n$. Consequently, the classical Shannon entropy of this distribution yields the value of the first term in Eq.~(\ref{Eq:Holevo}) irrespectively of the phase noise.  The second term in Eq.~(\ref{Eq:Holevo}) is simply given by the von Neumann entropy of any individual output state $S( \hat{\varrho}'_\varphi)$ which does not depend on the specific value of $\varphi$.

In Fig.~\subref*{Fig:holevo} we depict the Holevo quantity as a function of the mean photon number in the output pulse for several values of the dimensionless nonlinearity parameter $\varkappa$. In the case of a linear channel, when $\varkappa=0$, the Holevo quantity grows monotonically with the signal intensity. However, in the nonlinear case the Holevo quantity exhibits a maximum, which can be understood as an interplay between two effects. For low intensities the nonlinear phase noise has minute impact and increasing the mean photon number allows to encode more information in the phase variable thanks to the lower extent of the shot noise relative to the circumference of the ring. For high intensities the nonlinear phase noise makes coherent states diffused on the ring as seen in Fig.~\ref{Fig:state_evolution}(c,d) which effectively scrambles any information encoded in the phase variable.

Let us stress that the nonlinear phase noise is irreversible and it cannot be compensated at the channel output. Its origin can be understood intuitively as follows. The amplitude of the coherent pulse entering the nonlinear medium exhibits vacuum-level fluctuations. The nonlinearity deforms these fluctuations in a nontrivial manner, which for short distances can be visualised as squeezing one field quadrature at the cost of expanding the conjugate one \cite{Andersen2016}. In the lossless case variances of these quadratures saturate the Heisenberg uncertainty relation, which indicates that the state remains pure and the transformation is reversible. However, in the case of distributed attenuation the propagating field exchanges fluctuations with the environment. The contribution to field fluctuations acquired that way is given at the vacuum level, which introduces excess noise \cite{Vogel1993,Leonhardt1993,PhysRevA.55.3117} that makes the state no longer Heisenberg-limited. Because the effect of losses is to lower the mean photon number of the Poissonian statistics of the pulse and the Kerr nonlinearity affects only the off-diagonal elements of the density matrix in the Fock basis that characterise phase coherence, this mixedness occurs for the phase variable.

\section{Kerr squeezing in a lossy medium}
\label{Sec:squeezing}

The Kerr effect in optical fibres can be used to generate squeezed states of light which exhibit quadrature fluctuations reduced below the shot noise level
\cite{RosenbluhShelbyPRL1991,Schmitt1998,Werner1998,Fiorentino2002}. We will now use the effective description of self-phase modulation in a lossy nonlinear medium developed in preceding sections to analyse the the attainable degree of squeezing in the presence of distributed attenuation.

\begin{figure}[h!tp]
	\centering
	\includegraphics[width=.7\textwidth]{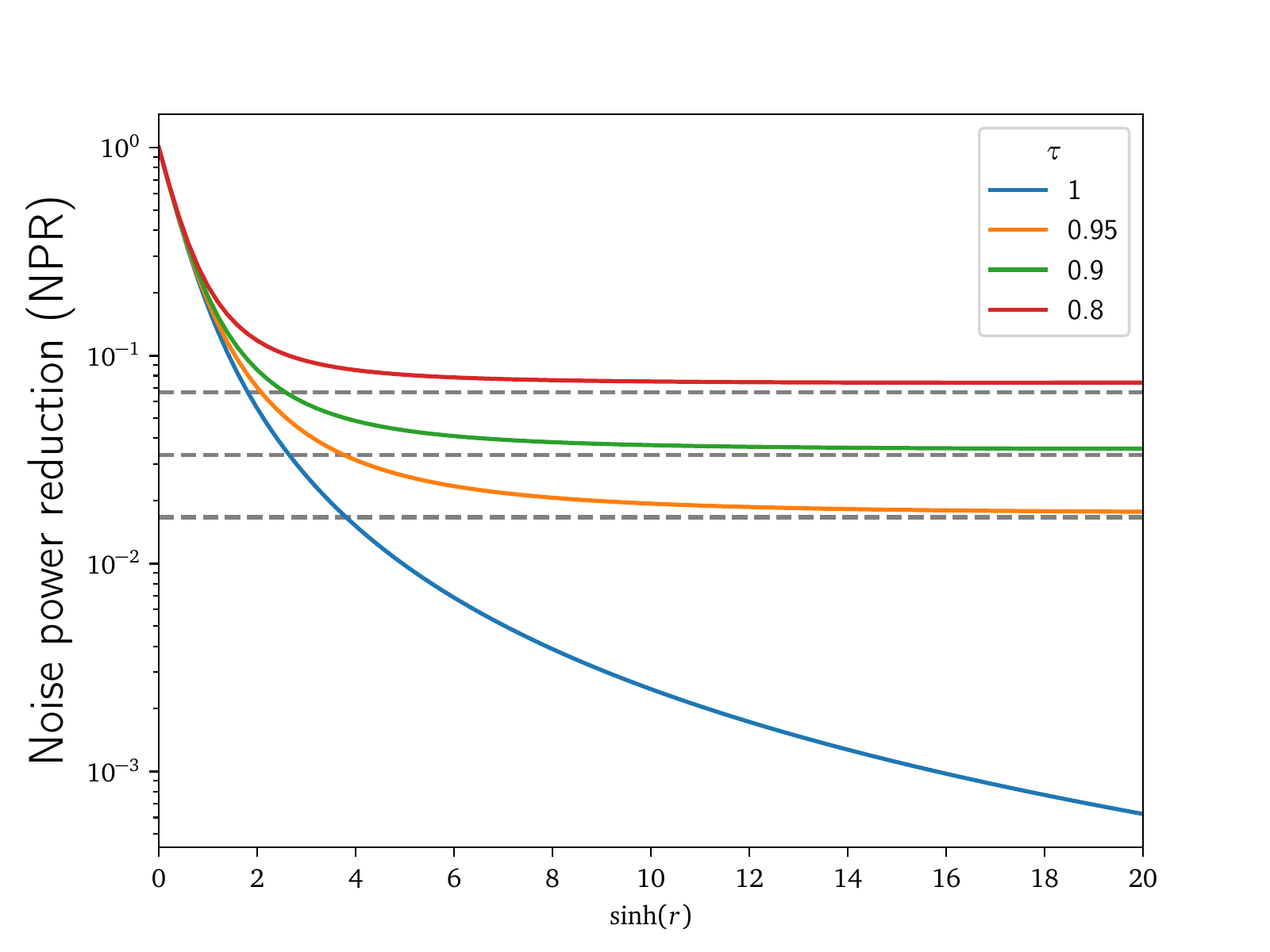}
	\caption{The attainable squeezing $\langle (\Delta\hat{x}_{\theta})^2\rangle$ for the input mean photon number $\bar n = 10^{8}$ and several different values of the transmission $\tau$ as a function of the dimensionless nonlinear coefficient $\varkappa$. For convenience, the abscissa is parameterised with $ \sinh r = 2 \varkappa (-\log\tau) \tau \bar{n} $. Dashed horizontal lines correspond to the estimate $(1-\tau)/3$. }
	\label{Fig:squeezing}
\end{figure}

In order to describe the output field emerging from the medium, we will take the phase diffused state evaluated using the Gaussian phase noise profile derived in Eq.~(\ref{Eq:GaussianNoiseDensityMatrix})
\begin{equation}
\hat{\varrho}'_G = \int_{-\infty}^{\infty} d \phi \frac{1}{\sqrt{2\pi \sigma^2}} e^{-\phi^2/2\sigma^2} \proj{e^{i\phi}\zeta}
\label{Eq:rho'squeezing}
\end{equation}
For notational simplicity we have chosen the phase distribution to be centred at $\phi=0$. We will also assume that the coherent amplitude $\zeta$ is real and positive, $\zeta=|\zeta|$. These two conditions can be satisfied by an appropriate choice of the input coherent state $\zeta_0$. The actual output state is obtained from $\hat{\varrho}'_G$ via the unitary Kerr transformation which yields $\exp(i \mu z \hat{n}^2) \hat{\varrho}'_G \exp(-i \mu z \hat{n}^2)$.
It will be most convenient to treat the unitary Kerr process $\exp(i \mu z \hat{n}^2)$ in the Heisenberg picture, in which the propagation equation for the annihilation operator reads
\begin{equation}
\frac{d\hat{a}}{dz} = - i \mu [\hat{n}^2 , \hat{a}(z)] = i \mu [ 2 \hat{a}^\dagger(z) \hat{a}(z)+1 ]\hat{a}(z) .
\end{equation}
Solving this equation allows us to express $\hat{a}(z)$ and $\hat{a}^\dagger(z)$ in terms of $\hat{a}(0)$ and $\hat{a}^\dagger(0)$. Expectation values involving the latter pair of operators should be taken over the ensemble of coherent states given in Eq.~(\ref{Eq:rho'squeezing}). For each individual pure state $\ket{e^{i\phi}\zeta}$ in the ensemble it will be convenient to use a substitution
\begin{equation}
\hat{a}(z) = e^{i \mu z (2\zeta^2 +1)} e^{ i \phi} [\zeta + \hat{b}(z)]
\label{Eq:SqueezingAnsatz}
\end{equation}
which shifts the reference frame to a location in which the initial mean field amplitude at $z=0$ is zeroed. This displacement maps the coherent state $\ket{e^{i\phi}\zeta}$ onto the vacuum state $\ket{0}$ which should be used to calculate expectation values of expressions involving operators
$\hat{b}(0)$ and $\hat{b}^\dagger(0)$. The overall phase factor $e^{i \mu z (2\zeta^2 +1) }$ has been introduced to simplify subsequent formulas.
The resulting propagation equation for the operator $\hat{b}(z)$ reads
\begin{equation}
\frac{d\hat{b}}{dz} =  2 i \mu \zeta^2 [\hat{b}(z) + \hat{b}^\dagger(z)],
\end{equation}
where on the right hand side we retained only terms linear in $\hat{b}(z)$ and $\hat{b}^\dagger(z)$. This is justified because $\hat{b}(z)$ and $\hat{b}^\dagger(z)$ describe only small fluctuations around the field amplitude which has been subtracted by the substitution (\ref{Eq:SqueezingAnsatz}). The solution of the linearized differential equation for $\hat{b}(z)$ is given by
\begin{equation}
\hat{b}(z) = (1+2 i \mu z \zeta^2 ) \hat{b}(0) + 2 i \mu z \zeta^2  \hat{b}^\dagger(0).
\label{Eq:b(t)solution}
\end{equation}
We will now introduce operators for two orthogonal quadratures of the output field defined as
\begin{eqnarray}
\hat{q} & = &  e^{-i \mu z (2\zeta^2 +1)} \hat{a}(z) + \text{h.c.} = e^{i\phi}[\zeta + \hat{b}(z)] + \text{h.c.}  ,\\
\hat{p} & = &  -i e^{-i \mu z (2\zeta^2 +1)} \hat{a}(z) + \text{h.c.} = -i  e^{i\phi} [\zeta + \hat{b}(z)] + \text{h.c.} ,
\end{eqnarray}
where h.c. denotes hermitian conjugated terms. Calculating statistical properties of output field quadratures involves performing two averages. The first one is the quantum mechanical expectation value for expressions involving $\hat{b}(z)$ and $\hat{b}^\dagger(z)$. These operators can be expressed using Eq.~(\ref{Eq:b(t)solution}) by $\hat{b}(0)$ and $\hat{b}^\dagger(0)$ which should be taken acting on the vacuum state.  The results are
\begin{equation}
\langle \hat{q} \rangle = 2 \zeta \langle \cos\phi \rangle, \qquad
\langle \hat{p} \rangle = 2 \zeta \langle \sin\phi \rangle
\end{equation}
in the first order and
\begin{eqnarray}
\langle  \hat{q}^2 \rangle & = & 1 + 4 \zeta^2   \langle \cos^2\phi \rangle - 4 \mu z \zeta^2 \langle \sin 2 \phi \rangle
+  (4\mu z \zeta^2)^2 \langle \sin^2\phi \rangle  \\
\langle  \hat{p}^2 \rangle & = & 1 + 4 \zeta^2   \langle \sin^2\phi \rangle + 4 \mu z \zeta^2 \langle \sin 2 \phi \rangle
+ (4 \mu z \zeta^2)^2 \langle \cos^2\phi \rangle
\end{eqnarray}
in the second order. The expectation value of the symmetically ordered product of the quadratures $\hat{q}$ and $\hat{p}$ reads
\begin{equation}
 \langle {\textstyle\frac{1}{2}} ( \hat{q}\hat{p} +  \hat{p} \hat{q} ) \rangle   =
 2\zeta^2 \langle \sin2 \phi \rangle + 4 \mu z \zeta^2 \langle \cos 2 \phi \rangle - 8 (\mu  t \zeta^2)^2 \langle \sin  2\phi \rangle
\end{equation}
The second average is over the random phase shift $\phi$ with a probability distribution
$\exp[-\phi^2/(2\sigma^2)]/\sqrt{2\pi\sigma^2}$. Because the distribution function is even we have $\langle \sin\phi \rangle = \langle \sin2\phi \rangle =0$ and the remaining averages can be determined from the expectation values
\begin{equation}
\langle \cos\phi \rangle = e^{-\sigma^2/2}, \qquad \langle \cos 2\phi \rangle = e^{-2\sigma^2}.
\end{equation}
It will be convenient to use in subsequent formulas a squeezing parameter $r$ defined through a relation $ \sinh r = 2 \mu z \zeta^2 $. Let us now consider quadrature fluctuations around the mean value given by $\Delta\hat{q} = \hat{q} - \langle \hat{q} \rangle$ and $\Delta\hat{p} = \hat{p} - \langle \hat{p} \rangle$. Their variances are explicitly given by
\begin{eqnarray}
\langle ( \Delta\hat{q})^2 \rangle & = & 1 +   2(1- e^{-\sigma^2})^2 \zeta^2 +  2(1-e^{-2\sigma^2}) \sinh^2 r   \label{Eq:Deltaq2}\\
\langle (\Delta\hat{p})^2 \rangle & = & 1 + 2(1-e^{-2\sigma^2}) \zeta^2 +   2(1+e^{-2\sigma^2}) \sinh^2 r
\end{eqnarray}
and the symmetrically ordered covariance reads
\begin{equation}
\langle {\textstyle\frac{1}{2}} ( \Delta\hat{q}\Delta\hat{p} +  \Delta\hat{p} \Delta\hat{q} ) \rangle
=   2 e^{-2\sigma^2} \sinh r.
\label{Eq:DeltaqDeltap}
\end{equation}
Let us now identify the angle $\theta$ for which a general quadrature given by $\hat{x}_\theta = \hat{q} \cos\theta + \hat{p}\sin\theta$ exhibits strongest squeezing.
It easy to verify that in the absence of phase noise, when $\sigma=0$, the minimum quadrature variance is obtained for $\tan\theta_0 = - e^{-r}$ and equals
$\langle (\Delta\hat{x}_{\theta_0})^2\rangle = e^{- 2 r}$.
When phase noise is included in the calculation, the optimal quadrature angle is given by
\begin{equation}
	\tan 2\theta = - \frac{\sinh r}{\sinh^2 r +  \zeta^2 (e^{\sigma^2}-1)}
\end{equation}
and the corresponding variance reads
\begin{eqnarray}
	\langle (\Delta\hat{x}_{\theta})^2\rangle &=& 1 + 2\zeta^2 (1-e^{-\sigma^2}) + 2\sinh^2 r \nonumber \\
	& & - 2 e^{-2\sigma^2} \sqrt{ \sinh^2 r +[ \sinh^2 r +\zeta^2 (e^{\sigma^2}-1)]^2} .
\end{eqnarray}
Let us recall that in our representation of lossy Kerr self-phase modulation the input coherent amplitude $\zeta_0$ enters the above expression through
$\zeta^2 = \tau \bar n$ and hence $ \sinh r = 2 \mu z \tau \bar n $, while $\sigma^2$ is given by Eq.~(\ref{Eq:sigmaapprox}).
In Fig.~\ref{Fig:squeezing} we depict attainable squeezing for several values of the transmission $\tau$ assuming $\bar n = 10^{8}$. The abcissa is parameterised with the nonlinear interaction strength $\mu$ defining the actual value of $\sinh r$. It is seen that for strong nonlinearity the squeezing is limited by the phase noise induced by the interplay between the self-phase modulation and losses.

The effective limit on attainable squeezing can be estimated using the following simple calculation. When exponents $e^{-\sigma^2}$ and $e^{-2\sigma^2}$ in Eqs.~(\ref{Eq:Deltaq2})-(\ref{Eq:DeltaqDeltap}) are expanded up to the first order in $\sigma^2$, the dominant effect of phase noise
is the addition of an excess noise term $(\Delta p)^2_{\text{exc}} = 4 \sigma^2 \zeta^2$ to the variance of $\hat{p}$. All other terms including $\sigma^2$ can be neglected, as they involve factors $\sinh r$ or $\sinh^2 r$ that are small compared to $\zeta^2$. This excess noise can be written as ${(\Delta p)^2_{\text{exc}} = 4 g(\tau) \sinh^2 r}$, where
\begin{equation}
g(\tau) = \frac{2-\tau-\tau(1-\log\tau)^2}{\tau(\log\tau)^2} \approx \frac{1}{3}(1-\tau)
\end{equation}
and the second approximate form is valid for the transmission $\tau$ close to one.
For low attenuation the optimal quadrature angle stays close to $\theta_0$ and hence the minimum quadrature variance can be approximated by
\begin{equation}
\langle (\Delta\hat{x}_{\theta_0})^2\rangle = e^{- 2 r} + (\Delta p)^2_{\text{exc}} \sin^2 \theta_0 =
e^{-2 r} + g(\tau) (1-e^{-2r})\tanh r.
\end{equation}
For substantial squeezing, when $r \gg 1$, we have $(1-e^{-2r})\tanh r \approx 1$ and consequently the second term simplifies to $g(\tau) \approx (1-\tau)/3$. This value defines the minimum quadrature variance in the presence of loss. It is noteworthy that it differs by a simple factor $1/3$ compared to the scenario when attenuation physically occurs after the squeezing transformation.
In realistic nonlinear media, one needs to include other effects, such as guided acoustic wave Brillouin scattering and Raman scaterring, which have more severe effect on the attainable squeezing \cite{CorneyPRA2008}.

\section{Conclusions}
\label{Sec:Conclusions}

We  have presented a theoretical model for the propagation of coherent states in a lossy medium with third-order nonlinearity in a dispersion-free setting. Within this model we have identified
nonlinear phase noise emerging from the interplay between optical loss and the Kerr nonlinearity. This excess noise is non-unitary and can not be compensated at the output. It has been shown that for a broad range of system parameters an effective description of this excess noise is given by a Gaussian distribution introducing also a nonlinear phase shift for the input state.  While the phase shift can be taken into account at the preparation stage, the Gaussian phase diffusion process is non-unitary and cannot be compensated.
Identification of the nonlinear phase noise allows us to decompose formally the description of propagation in a lossy nonlinear medium into a sequence of three distinct processes, namely standard attenuation followed by Gaussian phase diffusion and then unitary Kerr evolution. The presented decomposition is valid for coherent states. It can be generalised to arbitrary states of the electromagnetic field using the Glauber-Sudarshan $P$ quasiprobability distribution for the density operator \cite{SudarshanPRL1963,GlauberPRA1963}, but it should be kept in mind that the effective transformation of individual coherent states in this representation depends in a non-trivial manner on their amplitudes through the parameters of the Gaussian phase diffusion process.

We  also demonstrated that the nonlinear phase noise severely impairs the ability to encode classical information in the phase variable. Specifically, we considered phase shift keying using a continuous constellation of coherent states distributed on a ring with a fixed intensity. In the case of non-zero nonlinearity, the Holevo quantity, used as a measure of accessible information, exhibits a maximum for a finite input mean photon number. This is because for input coherent states with a sufficienty high mean photon number the phase becomes completely scrambled due to increasing diffusion. Consequently, when designing an optimal input ensemble of coherent states in both phase and intensity variables for transmission of classical information, the contribution from the high intensity region should approach that of a Poisson channel \cite{MartinezJOSAB2007}. It would be also interesting to investigate the ultimate quantum limit on the classical capacity of an optical channel with nonlinear phase noise, generalising results obtained recently for linear Gaussian models \cite{GiovannettiGarciaPatronNPH2014}. Finally, we investigated the interplay between nonlinearity and losses in the generation of squeezed states utilising a Kerr medium. We showed that the nonlinear phase noise can be used to estimate the attainable squeezing in the presence of distributed attenuation.

One should note that in the context of optical communication, nonlinear transformation of spontaneous emission contributed by signal amplification is known to introduce excess phase noise \cite{GordonMollenauerOL1990,DemirJLT2007}. This effect arises in the classical propagation model of electromagnetic fields. Our analysis incorporates fully quantum fluctuations contributed by both the input field and the loss mechanism, which are not taken into account in classical propagation equations. Hence, nonlinear phase noise discussed here appears even without signal amplification built into the optical link. Furthermore, when considering accessible information, we assumed that the reversible effects of the nonlinear propagation have been compensated by applying an appropriate unitary transformation at the channel output. Our quantum mechanical model can be extended to include amplifier noise using methods developed in \cite{PhysRevA.39.4628}.

\ack

We acknowledge insightful discussions with Ivan
H. Deutsch, R. Garc\'{\i}a-Patr\'{o}n, and Ch. Marquardt.
This work is part of the project ``Quantum Optical
Communication Systems'' carried out within the TEAM
programme of the Foundation for Polish Science co-financed by the European Union under the European
Regional Development Fund.

\appendix

\section{Solution of the Master equation}
\label{Sec:deriv}

In this Appendix we present a detailed calculation of the factor $c_j(z)$ used in the ansatz in Eq.~(\ref{Eq:ansatz}).
With the transformed operator $\hat{a}'(z) = \exp(-i \mu z \hat{n}^2) \hat{a} \exp(i \mu z \hat{n}^2)$ and the number operator $\hat n$ the Master equation
\begin{equation}
	\frac{d\hat{\varrho}'}{dz} = \frac{\alpha}{2} \left( 2 \hat{a}'(z) \hat{\varrho}'(z)
	[\hat{a}'(z)]^\dagger - \hat{n} \hat{\varrho}'(z) - \hat{\varrho}'(z) \hat{n} \right)
\end{equation}
can be reduced to a differential equation for the matrix elements of the density matrix which then reads
\begin{equation}
	\frac{d\varrho'_{m,n}}{dz} = \frac{\alpha}{2} \left [ 2 \rho_{m+1,n+1}'(z) \sqrt{(m+1)(n+1)} e^{2i \mu z (m-n)} - \rho_{m,n}'(z) (m+n) \right] .
	\label{Eq:de_ele_app}
\end{equation}
Inserting the ansatz given in Eq.~(\ref{Eq:ansatz})
into Eq.~(\ref{Eq:de_ele_app}) yields a differential equation for the factor $c_{j}(z)$ with $j=m-n$ that can be written down as
\begin{equation}
	\frac{d c_{j}}{d z}  = \alpha \bar n e^{-\alpha z} e^{2i \mu z j} c_{j}(z) .
\end{equation}
This first order differential equation is solved by standard separation of variables which yields
\begin{equation}
	c_{j}(z) = \exp\left( - \frac{\bar n \alpha}{\alpha - 2i \mu j } e^{-\alpha z} e^{2i \mu z j} + A_0 \right)
	= \exp\left( - \frac{\bar n \tau}{1- 2i \varkappa j } \tau^{-2i \varkappa j} + A_0 \right)
\end{equation}
where $A_0$ is the integration constant and in the last step we used $\tau = e^{-\alpha z}$, $\varkappa = \mu/\alpha$ and $e^{2i \mu z j} = e^{-2i \varkappa j\log \tau} = \tau^{-2i \varkappa j}$.
Inserting this solution into the ansatz~(\ref{Eq:ansatz}) yields
\begin{equation}
	\varrho'_{mn}(z) =  \frac{(\sqrt{\tau}\zeta_0)^m (\sqrt{\tau} \zeta_0^\ast)^n}{\sqrt{m! n!}}
\exp\left( - \frac{\bar n \tau^{1-2i \varkappa (m-n)}}{1 - 2i \varkappa (m-n) } + A_0 \right).
\end{equation}
The constant $A_0$ can be obtained from the initial condition
$
\varrho'_{mn}(0) = e^{-\bar n} {\zeta_0^m (\zeta_0^\ast)^n}/{\sqrt{m! n!}}  ,
$
which yields Eq.~(\ref{Eq:density_elements}) with the function $f_\tau(\varkappa)$ given by Eq.~(\ref{Eq:f}).

\section*{References}

\bibliographystyle{iopart-num}

\bibliography{lit}

\end{document}